\documentclass[a4paper,11pt]{article}

\usepackage{graphicx,lscape,relsize,epic}

\def\dfrac#1#2{{\displaystyle{#1\over#2}}}
\newcommand{\arcsinh}{\mathop{\mathrm{arcsinh}}}
\begin{document}
\thispagestyle{empty}

\begin{center}
${}$
\vspace{3cm}

{\huge\textbf{Time delay in the Einstein-Straus solution}} \\

\vspace{2cm}

{\Large Kheir-Eddine Boudjemaa\footnote{also at Centre Universitaire de Khenchela, Algeria.}, Mourad Guenouche and
Sami R. Zouzou }

\vspace{1.5cm}

Laboratoire de Physique Th\'{e}orique,
D\'{e}partement de Physique, Facult\'{e} des Sciences,
Universit\'{e} Mentouri-Constantine, Algeria
\end{center}

boudjemaa\_kh2006@yahoo.fr

guenouchemourad@live.com

szouzou2000@yahoo.fr

\vspace{2cm}

{\large\textbf{Abstract}}

\vspace{0.5cm}

We compute the time delay of strong lensing in the framework of
the Einstein-Straus solution. We compare the theory to the
observational bound on the time delay of the lens SDSS J1004+4112.

\vspace{1cm}


\section{Introduction}
Before the work of Rindler and Ishak \cite{rindler-ishak}, the
general believe was that the deflexion angle of light passing near
an isolated static and spherically symmetric mass is independent
of the cosmological constant. This believe was based on the
argument that the cosmological constant disappears from the
geodesic equation for massless particles. In september 2007,
Rindler and Ishak \cite{rindler-ishak} corrected this believe.
They pointed out that it is not sufficient to consider the
geodesic equation but also the metric itself must be considered.
In the work of Rindler and Ishak the source emitting the light and
the Earth were supposed to be at rest with respect to the lens.
Also all the masses, including those of the Earth and of the
source, were neglected, except the mass of the lens. Since then,
there is a rich controversy about whether or not a cosmological
constant modifies the bending of light near an isolated spherical
mass. Sereno \cite{sereno1,Sereno2}, Sch\"{u}cker
\cite{Sch-Lambda-lens,sch-proceed}, Miraghaei and Nouri-Zonoz
\cite{Miraghaei}, Kantowski, Chen and Dai \cite{Kantowski},
confirm Rindler and Ishak's result, while Khriplovich and Pomeransky
\cite{Khriplovich}, Park \cite{park}, Gibbons, Warnick and Werner
\cite{Gibbons}, Simpson, Peacock and Heavens \cite{Simpson}
contradict Rindler and Ishak's findings.

Recently, Sch\"{u}cker \cite{Sch-strong} redid the calculations of
the bending of light by a spherically symmetric mass distribution,
which is taken to be a cluster of galaxies by relaxing all the
previous hypotheses except that of sphericity: the observer and
the source are allowed to move with respect to the cluster, the
masses of the other clusters are included in the form of a
homogeneous isotropic dust and the observer as well as the source
are taken comoving with respect to the dust. The appropriate
framework of this computation is the Einstein-Straus solution
\cite{Einstein-Straus,Schucking} that matches the Kottler solution
at the inside of the Sch\"{u}cking radius with the Friedmann
solution at the outside. Einstein and Straus' first motivation was
to explain why the cosmic expansion does not affect smaller length
scales such as planetary and atomic systems. Sch\"{u}cker's
calculations confirm previous calculations by Ishak, Rindler,
Dossett, Moldenhauer and Allison \cite{Ishak-Rind-Doss}: Taking
into account realistic cosmic velocities attenuates the effect of
the cosmological constant on the bending of light without however
cancelling it. Although there has never been a claim that the time
delay was independent of a cosmological constant, it is also
interesting to do the computation of the time delay in the
framework of the Einstein-Straus solution.

We will use the same units as in reference \cite{Sch-strong}: astrometers
(am), astroseconds (as), astrograms (ag)
\begin{equation}
\begin{array}{rclcl}
\mbox{am}&=&1.30\cdot10^{26}\mbox{ m}&=&4221\mbox{ Mpc},\\
\:\mbox{as}&=&4.34\cdot10^{17}\mbox{ s}&=&13.8\mbox{ Gyr},\\
\mbox{ag}&=&6.99\cdot10^{51}\mbox{ kg}&=&3.52\cdot10^{21}\mbox{ M}_\odot ,
\end{array}
\end{equation}
where M$_\odot $ denotes one solar mass. In these units
\begin{equation}
\mbox{c}=1\mbox{am as}^{-1},\: 8\pi G=1\mbox{
am}^3\mbox{as}^{-2}\mbox{as}^{-1},\: H_0=1\mbox{ as}^{-1},\:
\hbar=3.86\cdot 10^{-121} \mbox{am}^2 \mbox{ as}^{-1} \mbox{ag}.
\end{equation}
For spatially flat universes, to which we will restrict ourselves
in the following, we may set the scale factor today $a_0=1$.
\section{The Einstein-Straus solution with a cosmological constant}

We will consider hereafter the Einstein-Straus solution \cite{Einstein-Straus,Schucking}
generalized to include the cosmological constant \cite{Balbinot} but
restrict ourselves to spatially flat universes. We will need the
Jacobian of the transformation passing between the Friedmann and
Schwarzschild coordinates to calculate the geodesics of photons.
Let us quote the results obtained by Sch\"{u}cker \cite{Sch-strong}. Let
$(T,r,\theta, \varphi)$ and $(t, \chi, \theta, \varphi)$ stand
respectively for Kottler and Friedmann coordinates. The Kottler
metric

\begin{equation}
ds^{2}=B(r)dT^{2}-B(r)^{-1}dr^{2}-r^{2}d\Omega ^{2},
\end{equation}
with
\begin{equation} B(r)=1-\dfrac{2GM}{r}-\dfrac{\Lambda
}{3}r^{2},
\end{equation}
prevails inside a vacuole of radius $r_{Sch\ddot{u}}(T)$,
$r<r_{Sch\ddot{u}}$. The Friedmann spatially flat metric is given
by:
\begin{equation}
ds^{2}=dt^{2}-a(t)^{2}\left( d\chi ^{2}+\chi ^{2}d\Omega
^{2}\right),
\end{equation}
with the scale factor $a(t)$ determined by the first order
Friedmann equation
\begin{equation}
\dfrac{da}{dt}=\sqrt{A/a +\Lambda a^2/3}, \label{first-Friedmann}
\end{equation}
where
\begin{equation}
 A=a_0^{3}\rho _{dust0}/3=1-\dfrac{\Lambda}{3}, \label{A}
\end{equation}
prevails outside the vacuole $\chi \geq \chi _{Sch\ddot{u}}$. It
is worthwhile to notice that due to (\ref{first-Friedmann}), the
scale factor is strictly monotonic. The two solutions are glued
together at the constant Sch\"{u}cking radius $\chi
_{Sch\ddot{u}}$
\begin{equation}
r_{Sch\ddot{u}}(T):=a(t)\chi _{Sch\ddot{u}}.
\end{equation}
By taking into account the fact that the central mass must be
equal to the dust density times the volume of the ball with
Sch\"{u}cking radius $r_{Sch\ddot{u}}$
\begin{eqnarray}
\hspace{1.2cm} A&\hphantom{:}=&\dfrac{2M}{8\pi\chi _{Sch\ddot{u}}^3}=\dfrac{2GM}{\chi _{Sch\ddot{u}}^3}  \\
B(r_{Sch\ddot{u}})&=:& B_{Sch\ddot{u}}=1-\dfrac{A}{a}\chi
_{Sch\ddot{u}}^{2}-\dfrac{\Lambda}{3}a^{2}\chi _{Sch\ddot{u}}^{2}.
\end{eqnarray}
It is also useful to introduce $C_{Sch\ddot{u}}$
\begin{equation}
C_{Sch\ddot{u}}:=\sqrt{1-B_{Sch\ddot{u}}}.
\end{equation}
Sch\"{u}cker \cite{Sch-strong} computed the Jacobian of the coordinate
transformation at the Sch\"{u}cking radius and also the inverse of
the Jacobian (corresponding to the Jacobian of the inverse
coordinate transformation $(t,\chi)\rightarrow (T,r)$) with the
results
\begin{equation}
\begin{array}{rclcrcl}
  \left. \dfrac{\partial t}{\partial T}\right\vert _{Sch\ddot{u}}&=&1,\hspace{2cm}& & \left. \dfrac{\partial t}{\partial r}\right\vert _{Sch\ddot{u}}&=&-\dfrac{C_{Sch\ddot{u}}}{B_{Sch\ddot{u}}} \\
  &&&&&& \\
  \left. \dfrac{\partial \chi }{\partial T}\right\vert _{Sch\ddot{u}}&=&-\dfrac{C_{Sch\ddot{u}}}{a}, && \left. \dfrac{\partial \chi }{\partial r}\right\vert
  _{Sch\ddot{u}}&=&\dfrac{1}{aB_{Sch\ddot{u}}},
\end{array}\label{Jacob}
\end{equation}
and
\begin{equation}
\begin{array}{rclcrcl}
  \left. \dfrac{\partial T}{\partial t}\right\vert_{Sch\ddot{u}}&=&\dfrac{1}{B_{Sch\ddot{u}}},\hspace{1.2cm}& & \left. \dfrac{\partial T}{\partial \chi }\right\vert _{Sch\ddot{u}}&=&a\dfrac{C_{Sch\ddot{u}}}{B_{Sch\ddot{u}}}, \\
  &&&&&& \\
  \left. \dfrac{\partial r}{\partial t}\right\vert _{Sch\ddot{u}}&=&C_{Sch\ddot{u}}, && \left. \dfrac{\partial r}{\partial \chi }\right\vert _{Sch\ddot{u}}&=&a.
\end{array} \label{Inv-Jacob}
\end{equation}
In the following we also need to pass between Kottler time $T$ and
Friedmann time $t$ at the Sch\"{u}cking radius. To this end we
will use the result also obtained by Sch\"{u}cker \cite{Sch-strong}
\begin{equation}
\left. \dfrac{dt}{dT}\right\vert _{Sch\ddot{u}}=B_{Sch\ddot{u}}.
\label{Kott-Fried-times}
\end{equation}
\section{Integrating the geodesics of light}
We have the following situation: a first photon is emitted by the
source, a quasar, at a time $t_S^\prime$ and follows an upper
straight line trajectory until its arrival at a time
$t_{Sch\ddot{u}S}^\prime$ on the Sch\"{u}cking sphere in the half
space containing the source.  It is then bent inside the
Sch\"{u}cking radius, until it emerges from the Sch\"{u}cking
sphere in the half space containing the Earth at time
$t_{Sch\ddot{u}E}^\prime$, then follows a straight line until its
arrival on Earth at time $t_0^\prime=0$. A second photon is
emitted by the quasar at a time $t_S$ follows a lower straight
line trajectory, arrives on the Sch\"{u}cking sphere in the half
space containing the source at time $t_{Sch\ddot{u}S}$, it is then
bent inside the Sch\"{u}cking sphere and emerges from the
Sch\"{u}cking sphere in the half space containing the Earth at
time $t_{Sch\ddot{u}E}$, follows again a straight line until its
receipt on Earth at the same time as the first photon
$t_0=t_0^\prime=0$. We will here be interested in the computation
of the time delay $t_S-t_S^\prime$.

\setlength{\unitlength}{1cm}
\hspace{-.5cm}
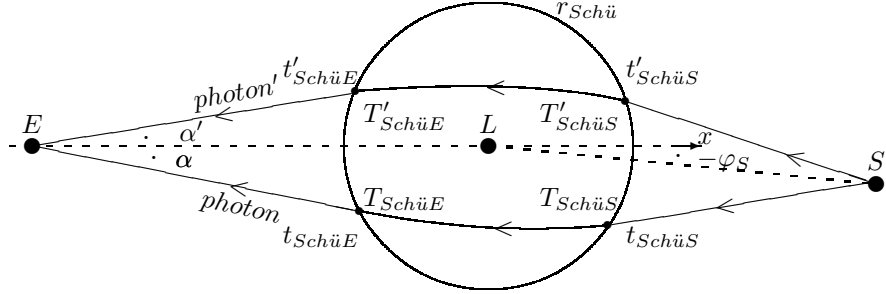
\begin{figure}[h]
\begin{picture}(6, 5)(-4,-1)
  \qbezier(4.9,2.0)(4.9,2.787)(4.3435,3.3435)
  \qbezier(4.3435,3.3435)(3.787,3.9)(3.0,3.9)
  \qbezier(3.0,3.9)(2.213,3.9)(1.6565,3.3435)
  \qbezier(1.6565,3.3435)(1.1,2.787)(1.1,2.0)
  \qbezier(1.1,2.0)(1.1,1.213)(1.6565,0.6565)
  \qbezier(1.6565,0.6565)(2.213,0.1)(3.0,0.1)
  \qbezier(3.0,0.1)(3.787,0.1)(4.3435,0.6565)
  \qbezier(4.3435,0.6565)(4.9,1.213)(4.9,2.0)
  \put(-3,2){\line(6,1){4.2}}
  \put(-3,2){\line(5,-1){4.35}}
  \put(4.8,2.6){\line(3,-1){3.3}}
  \put(4.6,0.95){\line(6,1){3.5}}
  {\small
  \put(0.9,2.4){\makebox(2,0){$T_{Sch\ddot{u}E}^\prime$}}
  \put(0.9,1.3){\makebox(2,0){$T_{Sch\ddot{u}E}$}}
  \put(3.2,1.3){\makebox(2,0){$T_{Sch\ddot{u}S}$}}
  \put(3.2,2.4){\makebox(2,0){$T_{Sch\ddot{u}S}^\prime$}}
  \put(-0.2,3.0){\makebox(2,0){$t_{Sch\ddot{u}E}^\prime$}}
  \put(-0.2,0.8){\makebox(2,0){$t_{Sch\ddot{u}E}$}}
  \put(4.3,0.8){\makebox(2,0){$t_{Sch\ddot{u}S}$}}
  \put(4.3,3.0){\makebox(2,0){$t_{Sch\ddot{u}S}^\prime$}}
  \put(-1.9,2.2){\makebox(2,0){$\alpha^\prime$}}
  \put(-2,1.8){\makebox(2,0){$\alpha$}}
  \put(5.1,1.8){\makebox(2,0){$-\varphi_S$}}
  \put(-2,1.8){\makebox(2,0){$\alpha$}}
  \put(-4,2.3){\makebox(2,0){$E$}}
  \put(2,2.3){\makebox(2,0){$L$}}
  \put(7.1,1.8){\makebox(2,0){$S$}}
  \put(3.3,3.8){\makebox(2,0){$r_{Sch\ddot{u}}$}}
  \put(4.85,2.1){\makebox(2,0){$x$}}
  \put(-0.9,2.5){\rotatebox{10}{$photon^\prime$}}
  \put(-0.6,2.507){\rotatebox{280}{$\vee $}}
  \put(-0.8,1.2){\rotatebox{-12}{$photon$}}
  \put(-0.4,1.58){\rotatebox{260}{$\vee $}}
  \put(3,2.89){\rotatebox{270}{$\vee $}}
  \put(3.1,1.02){\rotatebox{270}{$\vee $}}
  \put(6.9,1.984){\rotatebox{254}{$\vee $}}
  \put(6,1.296){\rotatebox{277}{$\vee $}}
  }
  \qbezier(1.25,2.73)(3.5,2.9)(4.8,2.6)
  \qbezier(1.3,1.14)(3,0.8)(4.57,0.95)
  \put(4.5,1.87){\makebox(2,0){$.$}}
  \put(-2.4,1.85){\makebox(2,0){$.$}}
  \put(-2.5,2.12){\makebox(2,0){$.$}}
  \put(1.25,2.73){\circle*{.1}}
  \put(1.3,1.14){\circle*{.1}}
  \put(4.8,2.6){\circle*{.1}}
  \put(4.57,0.95){\circle*{.1}}
  \put(-3,2){\circle*{.2}}
  \put(3,2){\circle*{.2}}
  \put(8.1,1.5){\circle*{.2}}
  \put(5.5,2){\vector(1,0){0.3}}
  \dashline{0.1}(-3.3,2)(5.5,2)
  \dashline{0.1}(3,2)(8,1.5)
  \end{picture}\caption{Two light rays emitted by a source $S$, bent inside the Sch\"{u}cking sphere and finally received at Earth
  $E$. The travel times of the two photons differ, giving rise to a time delay.}
\end{figure}


Let us first integrate the first order Friedmann solution
(\ref{first-Friedmann}) for the scale factor $a(t)$ in the spatially
flat case with cosmological constant $\Lambda$ and dust density $\rho_{dust0}=(3-\Lambda)$
(we will take the experimentally favored value $\Lambda=0.77\cdot
3$ am$^{-2}\pm 20\%$) with final condition $a(0)=1$. One may show
that (\ref{first-Friedmann}) is equivalent to the Friedmann second
order equation for $a(t)$
\begin{equation}
\dfrac{2}{a} \dfrac{d^2a}{dt^2}+\dfrac{1}{a^2}\left(\dfrac{da}{dt}
\right)^2=\Lambda, \label{second-Friedmann}
\end{equation}
with final conditions
\begin{equation}
\left.\dfrac{da}{dt}\right\vert_{t=0}=1,\qquad a(0)=1.
\end{equation}
The integration of (\ref{first-Friedmann}) or (\ref{second-Friedmann})
may be done analytically with the result
\begin{equation}
\hspace{-1cm}a(t)=\left(\dfrac{1-\Lambda/3}{\Lambda/3}\right)^{1/3}\left[\sinh^{2}\left(
\dfrac{3}{2}\sqrt{\dfrac{\Lambda}{3}}t+
\arcsinh\left(\dfrac{\Lambda/3}{1-\Lambda/3}\right)\right)\right]^{1/3}.
\label{a(t)}
\end{equation}
We will also need to solve the equation
\begin{equation}
\dfrac{d\chi}{dt}=-\dfrac{1}{a}\label{chi(t)}
\end{equation}
for various final conditions, with $\chi$ having the meaning of a
geodesic distance, not to be confused with a luminosity distance.
Since these are the final conditions, at the arrival on Earth,
which are known, we will proceed backwards in time in three steps:
we will determinate $t_{Sch\ddot{u}E}^\prime$ and
$t_{Sch\ddot{u}E}$, then $t_{Sch\ddot{u}S}^\prime$ and
$t_{Sch\ddot{u}S}$, and finally $t_S^\prime$ and $t_S$.
\begin{itemize}
\item [step 1:] Determination of $t_{Sch\ddot{u}E}^\prime$ and
$t_{Sch\ddot{u}E}$.

Here we will be interested in the photon trajectory between the
Earth and the Sch\"{u}cking sphere with the Friedmann metric. The
non vanishing Christoffel symbols of the Friedmann metric in the
plane $\theta=\pi/2$ are given by
\begin{equation}
\begin{array}{lclllcl}
  \Gamma _{\chi \chi }^{t}&=&aa_{t},& \hspace{1cm}&  \Gamma _{\varphi \varphi }^{t}&=&aa_{t}\chi ^{2}  \\
  &&&&& \\
  \Gamma _{\varphi \varphi }^{\chi }&=&-\chi, && \Gamma _{t\chi }^{\chi }&=&a_{t}/a,\\
  &&&&& \\
  \Gamma _{t\varphi }^{\varphi }&=&a_{t}/a,&& \Gamma _{\chi \varphi }^{\varphi }&=&1/\chi ,
\end{array}
\end{equation}
and the geodesic equation reads
\begin{equation}
\begin{array}{rl}
\ddot{t}+aa_{t}\dot{\chi}^{2}+aa_{t}\chi ^{2}\dot{\varphi}^{2}=&0,\\
&\\
\ddot{\chi}+2\dfrac{a_{t}}{a}\dot{t}\dot{\chi}-\chi \dot{\varphi}^{2}=&0,\\
&\\
\ddot{\varphi}+2\dfrac{a_{t}}{a}\dot{t}\dot{\varphi}+\dfrac{2}{\chi
}\dot{\chi}\dot{\varphi}=&0,
\end{array}
\end{equation}

with final conditions at $p=0$
\begin{equation}
\begin{array}{rlcrlcrl}
t~=&0,&\hspace{1cm}&\chi~=&\chi_E,&\hspace{1cm}&\varphi ~=&\pi\\
&&&&&&& \\
\dot{t}~=&1,&& \dot{\chi}~=&\cos{\alpha^\prime},
&&\dot{\varphi}~=&\dfrac{\sin{\alpha^\prime}}{\chi_E},
\end{array}
\end{equation}
for the upper trajectory photon, where we use the fact that the
physical angle $\alpha^\prime$ coincides with the coordinate
angle $\arctan(\left\vert \chi \dot{\varphi}/\dot{\chi}
\right\vert )$.

For the lower trajectory photon, the final conditions
differ
\begin{equation}
\begin{array}{rlcrlcrl}
t~=&0,&\hspace{1cm}&\chi~=&\chi_E,&\hspace{1cm}&\varphi ~=&-\pi\\
&&&&&&& \\
\dot{t}~=&1,&& \dot{\chi}~=&\cos{\alpha},
&&\dot{\varphi}~=&-\dfrac{\sin{\alpha}}{\chi_E}.
\end{array}
\end{equation}
The solution of the geodesic equation is
\begin{equation}
\begin{array}{rlcrlcrl}
\dot{t}~=&\dfrac{1}{a},&\hspace{0.3cm}&\dfrac{\chi^\prime_p}{\chi}~=&\sin{(\varphi-\alpha^\prime)},
&\hspace{0.4cm}&\dot{\varphi} ~=&\dfrac{\chi^\prime_p}{a^2\chi^2},
\end{array}
\end{equation}
where $\chi^\prime_p$, the would be peri-lens, is given by
\begin{equation}
\chi^\prime_p=\chi_E\sin{\alpha^\prime},
\end{equation}
for the upper trajectory photon and
\begin{equation}
\begin{array}{rlcrlcrl}
\dot{t}~=&\dfrac{1}{a},&\hspace{0.3cm}&\dfrac{\chi_p}{\chi}~=&\!\!-\sin{(\varphi+\alpha)},
&\hspace{0.3cm}&\dot{\varphi} ~=&\!\!-\dfrac{\chi_p}{a^2\chi^2},
\end{array}
\end{equation}
where $\chi_p$, the would be peri-lens, is given by
\begin{equation}
\chi_p=\chi_E\sin{\alpha},
\end{equation}
for the lower trajectory photon.

The polar angles $\varphi _{Sch\ddot{u}E}^{\prime }$ and $\varphi
_{Sch\ddot{u}E}$ at which the lower and the upper trajectory
photons emerge from the Sch\"{u}cking sphere are given
respectively by
\begin{equation}
\varphi _{Sch\ddot{u}E}^{\prime }=\pi-\arcsin\left( \dfrac{\chi_p^\prime}{\chi _{Sch\ddot{u}}} \right)+\alpha^\prime,
\end{equation}
\begin{equation}
\varphi _{Sch\ddot{u}E}=-\pi+\arcsin\left( \dfrac{\chi_p}{\chi _{Sch\ddot{u}}} \right)-\alpha,
\end{equation}
where $\chi _{Sch\ddot{u}}$, the Sch\"{u}cking radius, is given by
\begin{equation}
\chi _{Sch\ddot{u}}=\left( \dfrac{M}{4\pi(1-\Lambda/3)}\right)^{1/3},
\end{equation}

where we make use of (\ref{A}) together with the fact that we work
in system of units such that $A=1-\Lambda/3$. It is easy to see
using elementary Euclidean geometry that the geodesic distance
$\chi_{Sch\ddot{u}E,E}^{\prime }$ between the upper trajectory
photon when it emerges from the Sch\"{u}cking sphere and the Earth
is given by
\begin{equation}
\chi _{Sch\ddot{u}E,E}^{\prime }=\sqrt{\chi _{E}^{2}+\chi
_{Sch\ddot{u}}^{2}+2\,\chi _{E}\chi _{Sch\ddot{u}}\cos \varphi
_{Sch\ddot{u}E}^{\prime }},\label{chiEprime}
\end{equation}
where $\chi_E$ is the Earth-lens geodesic distance.

In the same way the geodesic distance $\chi_{Sch\ddot{u}E,E}$,
between the lower trajectory photon when it emerges from the
Sch\"{u}cking sphere and the Earth is given by an analogous
expression
\begin{equation}
\chi _{Sch\ddot{u}E,E}=\sqrt{\chi _{E}^{2}+\chi
_{Sch\ddot{u}}^{2}+2\,\chi _{E}\chi _{Sch\ddot{u}}\cos \varphi
_{Sch\ddot{u}E}}, \label{chiE}
\end{equation}

The Earth-Lens geodesic distance $\chi_E$ may be deduced from the knowledge of the redshift $z_E$ according to the scheme
\begin{equation}
z_E \:\rightarrow \: a_E=\dfrac{1}{1+z_E}\:\rightarrow \:
\widetilde{t}_E:= \widetilde{t}(a_E)  \:\rightarrow \: \chi(
\widetilde{t}_E)=:\chi_E,
\end{equation}
where $ \widetilde{t}(a)$ denotes the inverse of the scale factor
$a(t)$, where $\chi$ is the solution of the first order
differential equation (\ref{chi(t)}), subject to the initial
condition $\chi(0)=0$, meaning that the photon reaches the Earth
today at $t_0=0$. Moreover, since the scale factor $a$ is strictly
positif, $\chi$ is a strictly decreasing function and thus
injective. Therefore, $\chi(t)$ can be inverted to give $t$ in
term of $\chi$.

The knowledge of $z_E$ allows to determine $a_E$, then one deduces
the corresponding time $\widetilde{t}_E$. Injecting in $\chi(t)$,
one finally deduces $\chi_E$. This is the same procedure that will
be used to deduce $\chi_S$
\begin{equation}
z_S \:\rightarrow \: a_S=\dfrac{1}{1+z_S}\:\rightarrow \:
\widetilde{t}_S:= \widetilde{t}(a_S)  \:\rightarrow \: \chi(
\widetilde{t}_S)=:\chi_S.
\end{equation}
On the other hand, the geodesic distance between the upper
trajectory photon and the Earth in the time interval from the
crossing of the Sch\"{u}cking sphere in the half space containing
the Earth  $t_{Sch\ddot{u}E}^{\prime }$ until its arrival at Earth
$t_0^{\prime }=0$ is governed by equation (\ref{chi(t)}) with the
final condition
\begin{equation}
\chi(0)=0 \label{chi0}
\end{equation}
meaning that at $t_0^{\prime }=0$ the photon reaches the Earth.
From (\ref{chi(t)}) and (\ref{chi0}), one deduces that
\begin{equation}
\int_{0}^{\chi _{Sch\ddot{u}E,E}^{\prime }}d\chi=-\int^{t_{Sch\ddot{u}E}^{\prime }}_0\frac{1}{a(t)}dt
\end{equation}
i.e.,
\begin{equation}
\chi _{Sch\ddot{u}E,E}^{\prime }=-\int^{t_{Sch\ddot{u}E}^{\prime
}}_0\frac{1}{a(t)}dt=\int_{t_{Sch\ddot{u}E}^{\prime
}}^0\frac{1}{a(t)}dt. \label{chiEprime1}
\end{equation}
Then, comparing with equation (\ref{chiEprime}), one gets
\begin{eqnarray}
\sqrt{\chi _{E}^{2}+\chi _{Sch\ddot{u}}^{2}+2\,\chi _{E}\chi
_{Sch\ddot{u}}\cos \varphi _{Sch\ddot{u}E}^\prime}
=\int_{t_{Sch\ddot{u}E}^{\prime
}}^0\frac{1}{a(t)}dt.\label{tprimeE}
\end{eqnarray}
Equation (\ref{tprimeE}) may then be used to deduce
$t_{Sch\ddot{u}E}^{\prime }$.

The geodesic distance between the lower trajectory photon and the
Earth in the time interval from the crossing of the Sch\"{u}cking
sphere in the half space containing Earth, $t_{Sch\ddot{u}E}$,
until its arrival at Earth $(t_0=t^\prime _0=0)$, is also governed
by (\ref{chi(t)}) with the final condition (\ref{chi0}). In an
analogous way, one gets
\begin{equation}
\chi
_{Sch\ddot{u}E,E}=\int^{0}_{t_{Sch\ddot{u}E}}\frac{1}{a(t)}dt=\int_{t_{Sch\ddot{u}E}}^0\frac{1}{a(t)}dt.
\label{chiE1}
\end{equation}
Then, comparison with equation (\ref{chiE}) gives
\begin{eqnarray}
\sqrt{\chi _{E}^{2}+\chi _{Sch\ddot{u}}^{2}+2\,\chi _{E}\chi
_{Sch\ddot{u}}\cos \varphi _{Sch\ddot{u}E}}
=\int_{t_{Sch\ddot{u}E}}^0\frac{1}{a(t)}dt, \label{tE}
\end{eqnarray}
which may serve to deduce $t_{Sch\ddot{u}E}$.

However, we find it more reliable to proceed in a different way:
we compute $t_{Sch\ddot{u}E}$ by difference with
$t_{Sch\ddot{u}E}^\prime$. Combining (\ref{tprimeE}) and (\ref{tE}),
one gets
\begin{equation}
\chi _{Sch\ddot{u}E,E}-\chi _{Sch\ddot{u}E,E}^{\prime
}=\int_{t_{Sch\ddot{u}E}}^0\frac{1}{a(t)}dt-\int_{t_{Sch\ddot{u}E}^{\prime
}}^0\frac{1}{a(t)}dt=\int_{t_{Sch\ddot{u}E}}^{t_{Sch\ddot{u}E}^{\prime
}}\frac{1}{a(t)}dt. \label{diff-Chi-int}
\end{equation}

Since $a(t)$ does vary significantly only on cosmological time
scales and since $\left\vert t_{Sch\ddot{u}E}^{\prime
}-t_{Sch\ddot{u}E}\right\vert$ is very much smaller than
cosmological time scales then
\begin{equation}
\int_{t_{Sch\ddot{u}E}}^{t_{Sch\ddot{u}E}^{\prime }}\frac{1}{a(t)}dt\simeq \dfrac{t_{Sch\ddot{u}E}^{\prime }-t_{Sch\ddot{u}E}}{a(t_{Sch\ddot{u}E}^{\prime })}.\label{int-tE-tEprime}
\end{equation}
On the other hand, combining (\ref{chiEprime}) and (\ref{chiE})
\begin{eqnarray}
 \chi _{Sch\ddot{u}E,E}-\chi _{Sch\ddot{u}E,E}^\prime=&&\sqrt{\chi _{E}^{2}+\chi _{Sch\ddot{u}}^{2}+2\chi _{E}\chi _{Sch\ddot{u}}\cos \varphi _{Sch\ddot{u}E}}\\
&-&\sqrt{\chi _{E}^{2}+\chi _{Sch\ddot{u}}^{2}+2\chi _{E}\chi _{Sch\ddot{u}}\cos \varphi _{Sch\ddot{u}E}^\prime}.
\end{eqnarray}
But
\begin{eqnarray}
\cos \varphi _{Sch\ddot{u}E}^{\prime }&=&-\cos \left( -\arcsin \left( \frac{\chi _{E}\sin \alpha ^{\prime }}{\chi _{Sch\ddot{u}}}\right) +\alpha ^{\prime }\right)\nonumber\\
&\simeq& -\cos \left( -\frac{\chi _{E}\sin \alpha ^{\prime }}{\chi _{Sch\ddot{u}}}+\alpha ^{\prime }\right) \nonumber\\
&\simeq& -1+\frac{1}{2}\left( \frac{\chi _{E}-\chi _{Sch\ddot{u}}}{\chi _{Sch\ddot{u}}}\right) ^{2}\alpha ^{\prime 2},
\end{eqnarray}
and similarly
\begin{equation}
\cos \varphi _{Sch\ddot{u}E}\simeq -1+\frac{1}{2}\left( \frac{\chi _{E}-\chi _{Sch\ddot{u}}}{\chi _{Sch\ddot{u}}}\right) ^{2}\alpha ^{ 2},
\end{equation}
and then one gets up to second order in the physical angles $\alpha$ and $\alpha^\prime$
\begin{equation}
 \chi _{Sch\ddot{u}E,E}-\chi _{Sch\ddot{u}E,E}^{\prime }\simeq
\left( \chi _{E}-\chi _{Sch\ddot{u}}\right) \dfrac{\chi
_{E}}{2\chi _{Sch\ddot{u}}}\left( \alpha ^{2}-\alpha ^{\prime
2}\right) .\label{diff-Chi-racine}
\end{equation}
Combining (\ref{int-tE-tEprime}) and (\ref{diff-Chi-racine}), one
gets an approximate expression for $t_{Sch\ddot{u}E}^{\prime
}-t_{Sch\ddot{u}E}$
\begin{equation}
 t_{Sch\ddot{u}E}^{\prime }-t_{Sch\ddot{u}E}\simeq
a(t_{Sch\ddot{u}E}^{\prime })(\chi _{E}-\chi
_{Sch\ddot{u}})\dfrac{\chi _{E}}{2\chi _{Sch\ddot{u}}}\left(
\alpha ^{2}-\alpha ^{\prime 2}\right),\label{diff-t-app}
\end{equation}
which may be used to deduce $t_{Sch\ddot{u}E}$ if
$t_{Sch\ddot{u}E}^{\prime }$ has been determined from
(\ref{tprimeE}). Since $\alpha >\alpha ^\prime$, the lower
trajectory photon emerges from the Sch\"{u}cking sphere before the
upper trajectory photon.

The upper trajectory photon emerges from the Sch\"{u}cking sphere
with 4-velocity
\begin{equation}
 \dot{t}_{Sch\ddot{u}E}^{\prime }=\dfrac{1}{a_{Sch\ddot{u}E}^\prime},\: \dot{\chi}_{Sch\ddot{u}E}^{\prime }=-\dfrac{\cos{(\varphi_{Sch\ddot{u}E}^\prime-\alpha^\prime)}}{a_{Sch\ddot{u}E}^{\prime 2}},\: \dot{\varphi}_{Sch\ddot{u}E}^\prime=\dfrac{\chi _{p}^\prime}{a_{Sch\ddot{u}E}^{\prime 2}\chi_{Sch\ddot{u}}^{ 2}},
\end{equation}
where
\begin{equation}
a_{Sch\ddot{u}E}^\prime:= a(t_{Sch\ddot{u}E}^\prime),
\end{equation}
and the lower trajectory photon emerges from the Sch\"{u}cking
sphere with 4-velocity
\begin{equation}
 \dot{t}_{Sch\ddot{u}E}=\dfrac{1}{a_{Sch\ddot{u}E}},\: \dot{\chi}_{Sch\ddot{u}E}=-\dfrac{\cos{(\varphi_{Sch\ddot{u}E}+\alpha)}}{a_{Sch\ddot{u}E}^2},\: \dot{\varphi}_{Sch\ddot{u}E}=\dfrac{-\chi _{p}}{a_{Sch\ddot{u}E}^2\chi_{Sch\ddot{u}}^2},
\end{equation}
where
\begin{equation}
a_{Sch\ddot{u}E}:= a(t_{Sch\ddot{u}E}).
\end{equation}
Let $\gamma_{F}^\prime$, $\gamma_{F}$ be the smaller physical
angles between the un-oriented direction of the upper trajectory
photon and the direction towards the lens, and between the
un-oriented direction of the lower trajectory photon and the
direction towards the lens. We have
\begin{equation}
 \gamma_{F}^\prime = \arctan \left(\left\vert \chi_{Sch\ddot{u}}\dfrac{\dot{\varphi}_{Sch\ddot{u}E}^\prime}{\dot{\chi}_{Sch\ddot{u}E}^\prime}\right\vert %
\right)=\pi-\left(\varphi_{Sch\ddot{u}E}^\prime -\alpha^\prime\right)=\arcsin\left(\dfrac{\chi_{p}^\prime}{\chi_{Sch\ddot{u}}}\right)
\end{equation}
and
\begin{equation}
 \gamma_{F} = \arctan \left(\left\vert \chi_{Sch\ddot{u}}\dfrac{\dot{\varphi}_{Sch\ddot{u}E}} {\dot{\chi}_{Sch\ddot{u}E}} \right\vert\right) %
=\pi+\left(\varphi_{Sch\ddot{u}E} +\alpha\right)=\arcsin\left(\dfrac{\chi_{p}}{\chi_{Sch\ddot{u}}}\right).
\end{equation}

\item [step 2:] Determination of $t_{Sch\ddot{u}S}^\prime$ and
$t_{Sch\ddot{u}S}$.

We have first to translate the 4-velocities of the upper and lower
trajectories photons into the coordinates $(T,r,\varphi)$. They
will serve as final conditions for the geodesic equation inside
the Sch\"{u}cking sphere where prevails Kottler metric.
 Using the inverse Jacobian (\ref{Inv-Jacob}), one gets
\begin{equation}
\dot{r}_{Sch\ddot{u}E}^\prime=\dfrac{C_{Sch\ddot{u}E}^\prime-cos\left(\varphi_{Sch\ddot{u}E}^\prime
-\alpha^\prime \right)}{a_{Sch\ddot{u}E}^\prime}
\end{equation}
and
\begin{equation}
\dot{r}_{Sch\ddot{u}E}=\dfrac{C_{Sch\ddot{u}E}-cos\left(\varphi_{Sch\ddot{u}E}
+\alpha \right)}{a_{Sch\ddot{u}E}},
\end{equation}
with
\begin{equation}
C_{Sch\ddot{u}E}^\prime=C_{Sch\ddot{u}}(t_{Sch\ddot{u}E}^\prime)
\end{equation}
and
\begin{equation}
C_{Sch\ddot{u}E}=C_{Sch\ddot{u}}(t_{Sch\ddot{u}E}),
\end{equation}
where $C_{Sch\ddot{u}}(t)$ is given by
\begin{eqnarray}
C_{Sch\ddot{u}}(t)&=&\sqrt{1-B_{Sch\ddot{u}}(t)}=\sqrt{\dfrac{A}{a(t)}\chi_{Sch\ddot{u}}^2+ \dfrac{\Lambda}{3}a^2(t)\chi_{Sch\ddot{u}}^2}\\
&=&\chi_{Sch\ddot{u}}\sqrt{\dfrac{A}{a(t)}+ \dfrac{\Lambda}{3}a^2(t)}.
\end{eqnarray}

Let $\gamma_{K}^\prime$ and $\gamma_{K}$ denote respectively the
smaller coordinate angles between the un-oriented direction of the
upper trajectory photon and the direction towards the lens and
between the un-oriented direction of the lower trajectory photon
and the direction towards the lens. We have
\begin{equation}
\gamma_{K}^\prime=:\arctan\left( \left\vert r_{Sch\ddot{u}E}^\prime\dfrac{\dot{\varphi}_{Sch\ddot{u}E}^\prime} {\dot{r}_{Sch\ddot{u}E}^\prime} \right\vert\right)
=\arctan\left(\frac{\sin \gamma _F^\prime}{C_{Sch\ddot{u}E}^\prime+\cos \gamma _{F}^\prime}\right),
\end{equation}
and
\begin{equation}
\gamma_{K}=:\arctan \left(\left\vert r_{Sch\ddot{u}E}\dfrac{\dot{\varphi}_{Sch\ddot{u}E}} {\dot{r}_{Sch\ddot{u}E}} \right\vert\right)
=\arctan \left(\frac{\sin \gamma _F}{C_{Sch\ddot{u}E}+\cos \gamma _{F}}\right).
\end{equation}

Moreover, we have at our disposal an initial condition, \cite{Sch-strong},
which we may use to set
$T_{Sch\ddot{u}E}^\prime=t_{Sch\ddot{u}E}^\prime$. We have thus
specified the final conditions of the geodesic equation inside the
Sch\"{u}cking sphere.

Making use of the Christoffel symbols of the Kottler metric in the
equatorial plane $\theta=\pi/2$
\begin{equation}
 \Gamma _{Tr}^{T}=\dfrac{B^{\prime }}{2B},\quad\Gamma _{TT}^{r}=\dfrac{BB^{\prime }}{2},\quad\Gamma _{rr}^{r}=-\dfrac{B^{\prime }}{2B},\quad\Gamma _{\varphi \varphi }^{r}=-rB,\quad\Gamma _{r\varphi }^{\varphi }=\dfrac{1}{r},
\end{equation}
the geodesic equations then read
\begin{eqnarray}
\ddot{T}+\frac{B^{\prime }(r)}{B(r)}\dot{T}\dot{r}=0,\\
\ddot{r}+\frac{1}{2}B(r)B^{\prime }(r)\dot{T}^{2}-\frac{1}{2}\frac{B^{\prime }(r)}{B(r)}\dot{r}^{2}-r B(r)\,\dot{\varphi}^{2}=0,\\
\ddot{\varphi}+\frac{2}{r}\dot{r}\dot{\varphi}=0,
\end{eqnarray}
from which we deduce three first integrals
\begin{eqnarray}
\dot{T}=1/B(r), \label{fit}\\
\dot{\varphi}r^{2}=J,  \label{fifi}\\
\frac{\dot{r}^{2}}{B(r)}+\frac{J^{2}}{r^{2}}-\frac{1}{B(r)}=-E.
\label{fir}
\end{eqnarray}
Equation (\ref{fifi}) comes from invariance of the metric under
rotations and $J$ has the meaning of an angular momentum per unit
mass, (\ref{fir}) comes from invariance of the metric under time
translations and $E$ has the meaning of energy per unit mass. For the
photon $E=0$.

Eliminating the affine parameter between (\ref{fifi}) and (\ref{fir}),
one gets in the case of a photon
\begin{equation}
\frac{dr}{d\varphi }=\pm
r\sqrt{\frac{r^{2}}{J^{2}}-B}.\label{dr-dfi-kottler}
\end{equation}
At the peri-lens $r_p^\prime$ for the upper trajectory photon
$\left. \frac{dr}{d\varphi }\right\vert _{r_{p}^{\prime }}=0$,
from which one deduces the expression of $J$ in terms of
$r_p^\prime$
\begin{equation}
J=\frac{r_{p}^{\prime }}{\sqrt{B(r_{p}^{\prime
})}}.\label{J-prime}
\end{equation}
Similarly at the peri-lens $r_p$ for the lower trajectory photon
$\left. \frac{dr}{d\varphi }\right\vert _{r_{p}}=0$ and one
deduces  another expression for $J$ in terms of $r_p$
\begin{equation}
J=\frac{r_{p}}{\sqrt{B(r_{p})}}.\label{J}
\end{equation}
Replacing $J$ by the appropriate expression (\ref{J-prime}) or
(\ref{J}), one gets
\begin{equation}
\dfrac{d\varphi }{dr}=\pm \dfrac{1}{r\sqrt{r^{2}/r_{p}^{\prime
2}-1}}\left( 1-\dfrac{s}{r}-\dfrac{s}{r_{p}^{\prime
}}\dfrac{r}{r+r_{p}^{\prime }}\right) ^{-1/2}, \label{dfidr-prime}
 \label{dfi-dr-prime}
\end{equation}
valid for the upper trajectory photon and
\begin{equation}
\dfrac{d\varphi }{dr}=\pm
\dfrac{1}{r\sqrt{r^{2}/r_{p}^{2}-1}}\left(
1-\dfrac{s}{r}-\dfrac{s}{r_{p}}\dfrac{r}{r+r_{p}}\right) ^{-1/2}. \label{dfidr}
\label{dfi-dr}
\end{equation}
valid for the lower trajectory photon, where $s$ denotes the
Schwarzschild radius $s=2GM$.

It is worthwhile to notice that the cosmological constant has
disappeared from (\ref{dfi-dr-prime}) and also from (\ref{dfi-dr}).

Since $s/r_p^\prime \ll 1 $ and $s/r_p \ll 1 $, we will hereafter
only retain terms up to linear order in $s/r_p^\prime $ or in
$s/r_p$. In this approximation
\begin{equation}
r_{p}^{\prime }\simeq a(t_{Sch\ddot{u}E}^{\prime })\chi
_{Sch\ddot{u}}\sin \gamma _{K}^{\prime }-GM
\end{equation}
and
\begin{equation}
r_{p}\simeq a(t_{Sch\ddot{u}E})\chi _{Sch\ddot{u}}\sin \gamma
_{K}-GM.
\end{equation}
Eliminating now the affine parameter between (\ref{fit}) and
(\ref{fir}) and taking into account (\ref{J-prime}), one gets
\begin{equation}
\dfrac{dT}{dr}=\pm \dfrac{1}{B(r)\sqrt{1-\dfrac{r_{p}^{\prime
2}}{r^{2}}\dfrac{B(r)}{B(r_{p}^{\prime })}}}, \label{dT-dr-prime}
\end{equation}
for the upper trajectory photon and
\begin{equation}
\dfrac{dT}{dr}=\pm
\dfrac{1}{B(r)\sqrt{1-\dfrac{r_{p}^{2}}{r^{2}}\dfrac{B(r)}{B(r_{p})}}},\label{dT-dr}
\end{equation}
for the lower trajectory photon.

Let us denote $T_{Sch\ddot{u}S}^{\prime }$ and
$T_{Sch\ddot{u}E}^{\prime }$ the Kottler times at which the upper
trajectory photon penetrates inside and emerges from the vacuole
respectively and $t_{Sch\ddot{u}S}^{\prime }$ and
$t_{Sch\ddot{u}E}^{\prime }$ the corresponding Friedmann times.
From now on, we will denote the expression of $d\varphi/dr$ and
$dT/dr$ valid for the upper trajectory photon by
$d\varphi^\prime/dr$ and $dT^\prime/dr$ respectively and will
continu to denote the expressions valid for the lower trajectory
by $d\varphi/dr$ and $dT/dr$. We have
\begin{equation}
T_{Sch\ddot{u}E}^{\prime }-T_{Sch\ddot{u}S}^{\prime
}=\!\int_{r_{P}^{\prime }}^{r( t_{Sch\ddot{u}E}^\prime) }\left\vert
\frac{dT^{\prime }}{dr}\right\vert dr+\!\int_{r_{p}^\prime}^{r(
t_{Sch\ddot{u}S}^\prime) }\left\vert \frac{dT^{\prime
}}{dr}\right\vert dr, \label{TEprime-TSprime}
\end{equation}
with $r(t)=a(t)\chi _{Sch\ddot{u}}$ and where $dT^{\prime }/dr$ is
given by (\ref{dT-dr-prime}). To obtain (\ref{TEprime-TSprime}), we
have used the fact that $T^\prime$ decreases when $r$ increases
from $r( t_{Sch\ddot{u}S}^\prime)$ to $r_p^\prime$ as well as when
$r$ increases from $r_p^\prime$ to $r( t_{Sch\ddot{u}E}^\prime)$.

Using the relation (\ref{Kott-Fried-times}) relating the Kottler
time $T$ and the Friedmann time $t$, one gets
\begin{equation}
T_{Sch\ddot{u}S}^\prime=T_{Sch\ddot{u}E}^\prime-\int^{t_{Sch\ddot{u}E}^\prime}_{t_{Sch\ddot{u}S}^\prime}
\dfrac{dt}{B_{Sch\ddot{u}}(t)}. \label{KotTsch-prime}
\end{equation}
Moreover, we have at our disposal a free initial condition
\cite{{Sch-strong}} which we may use to set $ t_{Sch\ddot{u}E}^{\prime }=
T_{Sch\ddot{u}E}^{\prime }$ for instance.
Combining (\ref{TEprime-TSprime}) and (\ref{KotTsch-prime}), one gets
\begin{equation}
 \int_{r_{P}^{\prime }}^{r( t_{Sch\ddot{u}E}^{\prime }) } \left\vert \dfrac{dT^{\prime }}{dr}\right\vert dr+\int_{r_{P}^{\prime }}^{r( t_{Sch\ddot{u}S}^{\prime }) }\left\vert \dfrac{dT^{\prime }}{dr}\right\vert dr-\int_{t_{Sch\ddot{u}S}^{\prime }}^{t_{Sch\ddot{u}E}^{\prime }}\frac{dt}{B_{Sch\ddot{u}}(t)}=0, \label{KotTsch-prime-int}
\end{equation}
from which one can deduce $ t_{Sch\ddot{u}S}^{\prime }$. If one is
interested in $T_{Sch\ddot{u}S}^{\prime }$ one can use
(\ref{KotTsch-prime}) to obtain it.

Let us denote $T_{Sch\ddot{u}S}$ and $T_{Sch\ddot{u}E}$ the
Kottler times at which the lower trajectory photon penetrates into
and leaves the Sch\"{u}cking sphere respectively and by
$t_{Sch\ddot{u}S}$ and $t_{Sch\ddot{u}E}$ the corresponding
Friedmann times.

To compute $t_{Sch\ddot{u}S}$, one can proceed in a similar manner as for $ t_{Sch\ddot{u}S}^{\prime }$. We have the analogous of (\ref{TEprime-TSprime}), (\ref{KotTsch-prime}) and (\ref{KotTsch-prime-int})
\begin{eqnarray}
T_{Sch\ddot{u}E}-T_{Sch\ddot{u}S}&=&\!\!\int_{r_{P}}^{r( t_{Sch\ddot{u}E}) }\left\vert \frac{dT}{dr}\right\vert dr+\int_{r_{P}^\prime}^{r( t_{Sch\ddot{u}S}) }\left\vert \frac{dT}{dr}\right\vert dr, \label{TE-TS}\\
T_{Sch\ddot{u}S}=T_{Sch\ddot{u}E}&-&\!\!\int^{t_{Sch\ddot{u}E}}_{t_{Sch\ddot{u}S}} \dfrac{dt}{B_{Sch\ddot{u}}(t)}\label{KotTsch}
\end{eqnarray}
and
\begin{equation}
 \int_{r_{P}}^{r( t_{Sch\ddot{u}E}) } \left\vert \dfrac{dT}{dr}\right\vert dr+\int_{r_{P}}^{r( t_{Sch\ddot{u}S}) }\left\vert \dfrac{dT}{dr}\right\vert dr-\int_{t_{Sch\ddot{u}S}}^{t_{Sch\ddot{u}E}}\frac{dt}{B_{Sch\ddot{u}}(t)}=0,\label{KotTsch-int}
\end{equation}
from which one can deduce $ t_{Sch\ddot{u}S}$.

But as in the cas of $ t_{Sch\ddot{u}E}$, we prefer to proceed in
a different manner: determining $ t_{Sch\ddot{u}S}$ by
differences. Combining (\ref{TEprime-TSprime}) and (\ref{TE-TS}),
one obtains
\begin{eqnarray}
 \left( T_{Sch\ddot{u}E}^{\prime }-T_{Sch\ddot{u}S}^{\prime }\right) -\left( T_{Sch\ddot{u}E}-T_{Sch\ddot{u}S}\right) =\int_{r_{p}^{\prime }}^{r( t_{Sch\ddot{u}E}^{\prime }) }\left\vert \frac{dT^{\prime }}{dr}\right\vert dr \nonumber \\
\hspace{-1cm}+\int_{r_{p}^{\prime }}^{r( t_{Sch\ddot{u}S}^{\prime }) }\left\vert \frac{dT^{\prime }}{dr}\right\vert dr
 -\int_{r_{p}}^{r( t_{Sch\ddot{u}E}) }\left\vert \frac{dT}{dr}\right\vert dr-\int_{r_{p}}^{r( t_{Sch\ddot{u}S}) }\left\vert \frac{dT}{dr}\right\vert dr.\label{Delta-TE-TS-prime}
\end{eqnarray}
On the other hand, using (\ref{Kott-Fried-times}), we have
\begin{equation}
T_{Sch\ddot{u}E}^{\prime }-T_{Sch\ddot{u}E}=\int_{t_{Sch\ddot{u}E}}^{t_{Sch\ddot{u}E}^{\prime }}\dfrac{dt}{B_{Sch\ddot{u}}(t)}\sim \frac{t_{Sch\ddot{u}E}^{\prime }-t_{Sch\ddot{u}E}}{B_{Sch\ddot{u}}(t_{Sch\ddot{u}E}^{\prime })},\label{Kott-TEprime-TE}
\end{equation}
and
\begin{equation}
T_{Sch\ddot{u}S}^{\prime }-T_{Sch\ddot{u}S}=\int_{t_{Sch\ddot{u}S}}^{t_{Sch\ddot{u}S}^{\prime }}\dfrac{dt}{B_{Sch\ddot{u}}(t)}\sim \frac{t_{Sch\ddot{u}S}^{\prime }-t_{Sch\ddot{u}S}}{B_{Sch\ddot{u}}(t_{Sch\ddot{u}S}^{\prime })}, \label{Kott-TSprime-TS}
\end{equation}
where we have used the fact that $B_{Sch\ddot{u}}$ does not vary
appreciably on the time intervals
$[t_{Sch\ddot{u}E},t_{Sch\ddot{u}E}^{\prime }]$ and
$[t_{Sch\ddot{u}S},t_{Sch\ddot{u}S}^{\prime }]$ since these are
smaller than cosmological scales.

Substituting (\ref{Kott-TEprime-TE}) and (\ref{Kott-TSprime-TS}) into (\ref{Delta-TE-TS-prime}) one gets
\begin{eqnarray}
\frac{t_{Sch\ddot{u}E}^{\prime }-t_{Sch\ddot{u}E}}{B_{Sch\ddot{u}}(t_{Sch\ddot{u}E}^{\prime })} &-&\frac{t_{Sch\ddot{u}S}^{\prime }-t_{Sch\ddot{u}S}}{B_{Sch\ddot{u}}(t_{Sch\ddot{u}S}^{\prime })}\simeq\nonumber\\
&&\int_{r_{p}^{\prime }}^{r(t_{Sch\ddot{u}E}^{\prime })}\left\vert \frac{dT^{\prime }}{dr}\right\vert dr+\int_{r_{p}^{\prime }}^{r(t_{Sch\ddot{u}S}^{\prime })}\left\vert \frac{dT^{\prime }}{dr}\right\vert dr\nonumber\\
&-&\int_{r_{p}}^{r( t_{Sch\ddot{u}E}) }\left\vert \frac{dT}{dr}\right\vert dr-\int_{r_{p}}^{r( t_{Sch\ddot{u}S}) }\left\vert \frac{dT}{dr}\right\vert dr
\end{eqnarray}
\begin{eqnarray}
\hspace{2cm}\simeq&&\quad\int_{r_{p}^{\prime }}^{r(t_{Sch\ddot{u}E}^{\prime })}\left\vert \frac{dT^{\prime }}{dr}\right\vert dr-\int_{r_{p}}^{r(t_{Sch\ddot{u}E}^{\prime })}\left\vert \frac{dT}{dr}\right\vert dr\nonumber\\
&&+\int_{r_{p}^{\prime }}^{r(t_{Sch\ddot{u}S}^{\prime })}\left\vert \frac{dT^{\prime }}{dr}\right\vert dr-\int_{r_{p}}^{r(t_{Sch\ddot{u}S}^{\prime })}\left\vert \frac{dT}{dr}\right\vert dr\nonumber\\
&&-\int_{r(t_{Sch\ddot{u}E}^{\prime })}^{r(t_{Sch\ddot{u}E})}\left\vert \frac{dT}{dr}\right\vert dr-\int_{r(t_{Sch\ddot{u}S}^{\prime })}^{r(t_{Sch\ddot{u}S})}\left\vert \frac{dT}{dr}\right\vert dr.
\end{eqnarray}
But since we deal with smaller length and time scales than cosmological ones:
\begin{eqnarray}
\int_{r(t_{Sch\ddot{u}E}^{\prime })}^{r(t_{Sch\ddot{u}E})}\left\vert \frac{dT}{dr}\right\vert dr&\simeq &\left\vert \frac{dT}{dr}\right\vert _{r(t_{Sch\ddot{u}E}^{\prime })}\left[ a(t_{Sch\ddot{u}E})-a(t_{Sch\ddot{u}E}^{\prime })\right] \chi _{Sch\ddot{u}},\nonumber\\
 \int_{r(t_{Sch\ddot{u}S}^{\prime })}^{r(t_{Sch\ddot{u}S})}\left\vert \frac{dT}{dr}\right\vert dr&\simeq& \left\vert \frac{dT}{dr}\right\vert _{r(t_{Sch\ddot{u}S}^{\prime })}\left[ a(t_{Sch\ddot{u}S})-a(t_{Sch\ddot{u}S}^{\prime })\right] \chi _{Sch\ddot{u}}.\nonumber
\end{eqnarray}
Using equation (\ref{first-Friedmann}), one deduces
\begin{eqnarray}
 a(t_{Sch\ddot{u}E})-a(t_{Sch\ddot{u}E}^{\prime })&\simeq &\sqrt{\frac{A}{a(t_{Sch\ddot{u}E}^{\prime })}+\frac{\Lambda }{3}a^{2}(t_{Sch\ddot{u}E}^{\prime })}\left( t_{Sch\ddot{u}E}-t_{Sch\ddot{u}E}^{\prime }\right),\\
 a(t_{Sch\ddot{u}S})-a(t_{Sch\ddot{u}S}^{\prime })&\simeq& \sqrt{\frac{A}{a(t_{Sch\ddot{u}S}^{\prime })}+\frac{\Lambda }{3}a^{2}(t_{Sch\ddot{u}S}^{\prime })}\left( t_{Sch\ddot{u}S}-t_{Sch\ddot{u}S}^{\prime }\right),
\end{eqnarray}
with
\begin{equation}
A=\dfrac{1}{3}\rho_{dust,0} \, a_0^3=\dfrac{1}{3}\rho_{dust,0}.
\end{equation}
On the other hand, using the results of reference
\cite{Sch-strong}
\begin{eqnarray}
 &&\int_{r_{p}^{\prime }}^{r(t_{Sch\ddot{u}E}^{\prime })}\left\vert \frac{dT^{\prime }}{dr}\right\vert dr-\int_{r_{p}}^{r(t_{Sch\ddot{u}E}^{\prime })}\left\vert \frac{dT}{dr}\right\vert \simeq \frac{M}{8\pi }\left[ \frac{1}{2}\left( 1-\frac{r_{p}^{\prime 2}}{r_{p}^{2}}\right) \frac{8\pi r_{p}^{2}}{M r(t_{Sch\ddot{u}E}^{\prime })}\right.\nonumber\\
&&\left. -\frac{3}{2}\left( \frac{r_{p}^{2}}{r_{p}^{\prime
2}}-1\right) \frac{M}{8\pi r_{p}^{2}\sqrt{\Lambda /3}}\mbox{
arctanh} \left( \sqrt{\frac{\Lambda}{3}}r(t_{Sch\ddot{u}E}^{\prime
})\right) -2\ln \left( \frac{r_{p}^{\prime }}{r_{p}}\right)
\right].
\end{eqnarray}
It suffices to make the following replacements
\begin{equation}
x\rightarrow \frac{r_{p}^{\prime }}{r_{p}},~\epsilon _{T}\rightarrow \frac{r_{p}}{r(t_{Sch\ddot{u}E}^{\prime })},~\delta \rightarrow \frac{M}{8\pi r_{p}},~\lambda \rightarrow r_{p}\sqrt{\Lambda /3}
\end{equation}
in equation (23) of reference \cite{sch-zaimen}.

In the same manner
\begin{eqnarray}
 &&\int_{r_{p}^{\prime }}^{r(t_{Sch\ddot{u}S}^{\prime })}\left\vert \frac{dT^{\prime }}{dr}\right\vert dr-\int_{r_{p}}^{r(t_{Sch\ddot{u}S}^{\prime })}\left\vert \frac{dT}{dr}\right\vert \simeq \frac{M}{8\pi }\left[ \frac{1}{2}\left( 1-\frac{r_{p}^{\prime 2}}{r_{p}^{2}}\right) \frac{8\pi r_{p}^{2}}{M r(t_{Sch\ddot{u}S}^{\prime })}\right. \nonumber\\
 &&\left. -\frac{3}{2}\left( \frac{r_{p}^{2}}{r_{p}^{\prime
2}}-1\right) \frac{M}{8\pi r_{p}^{2}\sqrt{\Lambda /3}}\mbox {
arctanh} \left( \sqrt{\frac{\Lambda}{3}}r(t_{Sch\ddot{u}S}^{\prime
})\right) -2\ln \left( \frac{r_{p}^{\prime }}{r_{p}}\right)
\right].
\end{eqnarray}

Collecting together the previous results, one gets an analytical approximation for $t_{Sch\ddot{u}S}-t_{Sch\ddot{u}S}^{\prime }$
\begin{equation}
\begin{array}{ll}
&\hspace{-0.5cm}\mbox{\fontsize{8}{13}\selectfont $ \displaystyle{t_{Sch\ddot{u}S}-t_{Sch\ddot{u}S}^{\prime }\simeq \left\{\frac{M}{8\pi }\left[ \frac{1}{2}\left( 1-\frac{r_{p}^{\prime 2}}{r_{p}^{2}}\right) \frac{8\pi r_{p}^{2}}{M\chi _{Sch\ddot{u}}}\left( \frac{1}{a(t_{Sch\ddot{u}E}^{\prime })}+\frac{1}{a(t_{Sch\ddot{u}S}^{\prime })}\right) \right.\right.}$}
\vspace{0.2cm}\\
&\mbox{\fontsize{8}{13}\selectfont $ \displaystyle{\left.
-\frac{3}{2}\left( \frac{r_{p}^{2}}{r_{p}^{\prime 2}}-1\right)
\frac{M}{8\pi r_{p}^{2}\sqrt{\Lambda /3}}\left( \mbox { arctanh}
\left( \sqrt{\frac{\Lambda}{3}}r(t_{Sch\ddot{u}E}^{\prime
})\right) +\mbox { arctanh} \left( \sqrt{\frac{\Lambda}{3}}r(t_{Sch\ddot{u}S}^{\prime
})\right)
\right) \right.}$}
 \vspace{0.2cm}\\
&\mbox{\fontsize{8}{13}\selectfont $ \displaystyle{\left.-4\ln
\left( \frac{r_{p}^{\prime }}{r_{p}}\right) \right]}$}
\vspace{0.2cm}\\
&\mbox{\fontsize{8}{13}\selectfont $ \displaystyle{+\left[ \chi _{Sch\ddot{u}}\sqrt{\frac{A}{a(t_{Sch\ddot{u}E}^{\prime })}+\frac{\Lambda }{3}a^{2}(t_{Sch\ddot{u}E}^{\prime })}\left\vert \frac{dT}{dr}\right\vert _{r(t_{Sch\ddot{u}E}^{\prime })}-\frac{1}{B_{Sch\ddot{u}}(t_{Sch\ddot{u}E}^{\prime })}\right]}$}
 \vspace{0.2cm}\\
&\mbox{\fontsize{8}{13}\selectfont $ \displaystyle{\left.\times a(t_{Sch\ddot{u}E}^{\prime })\left( \chi _{E}-\chi _{Sch\ddot{u}}\right) \frac{\chi _{E}}{2\chi _{Sch\ddot{u}}}\left( \alpha ^{2}-\alpha ^{\prime 2}\right)\vphantom{\left( \frac{r_{p}^{2}}{r_{p}^{\prime 2}}-1\right)}\right\}}$}
\vspace{0.2cm} \\
 &\mbox{\fontsize{8}{13}\selectfont $ \displaystyle{\times \left( \frac{1}{B_{Sch\ddot{u}}(t_{Sch\ddot{u}S}^{\prime })}+\chi _{Sch\ddot{u}}\sqrt{\frac{A}{a(t_{Sch\ddot{u}S}^{\prime })}+\frac{\Lambda }{3}a^{2}(t_{Sch\ddot{u}S}^{\prime })}\left\vert \frac{dT}{dr}\right\vert _{r(t_{Sch\ddot{u}S}^{\prime })}\right) ^{-1}}$},
\end{array}
\end{equation}
where we have also used the expression (\ref{diff-t-app}) for
$(t_{Sch\ddot{u}E}^\prime-t_{Sch\ddot{u}E})$. Then the knowledge
of $t_{Sch\ddot{u}S}^\prime$ allows one to deduce
$t_{Sch\ddot{u}S}$.

Let us now determine in turn the polar angles
$\varphi_{Sch\ddot{u}S}^\prime$ and $\varphi_{Sch\ddot{u}S}$ at
which the upper and lower trajectories photons penetrate inside
the Sch\"{u}cking sphere. Since the angle $\varphi^\prime$
increases all the way from $r_{Sch\ddot{u}S}^\prime$ to
$r_p^{\prime}$ and from $r_p^{\prime}$ to
$r_{Sch\ddot{u}E}^\prime$, one gets
\begin{equation}
 \varphi _{Sch\ddot{u}S}^{\prime }=\varphi _{Sch\ddot{u}E}^{\prime }-\int_{r_{p}^{\prime }}^{r( t_{Sch\ddot{u}S}^{\prime }) }\left\vert \frac{d\varphi ^{\prime }}{dr}\right\vert dr-\int_{r_{p}^{\prime }}^{r( t_{Sch\ddot{u}E}^{\prime }) }\left\vert \frac{d\varphi ^{\prime }}{dr}\right\vert dr.
\end{equation}
To the linear order in the ratio $s/r_p^\prime$, Schwarzschild
radius $s=2GM$ divided by peri-lens $r_p^\prime$, one gets
\cite{Sch-strong}
\begin{eqnarray}
\varphi _{Sch\ddot{u}S}^{\prime }\sim& \varphi _{Sch\ddot{u}E}^{\prime }-\pi +\arcsin \left(\dfrac{r_{p}^{\prime }}{r_{Sch\ddot{u}E}^{\prime }}\right)+\arcsin \left(\dfrac{r_{p}^{\prime }}{r_{Sch\ddot{u}S}^{\prime }}\right)\nonumber\\
&-\dfrac{1}{2}\dfrac{s}{r_{Sch\ddot{u}E}^{\prime}}\sqrt{\dfrac{r_{Sch\ddot{u}E}^{\prime 2}}{r_{p}^{\prime 2}}-1}-\frac{1}{2}\frac{s}{r_{Sch\ddot{u}S}^{\prime }}\sqrt{\dfrac{r_{Sch\ddot{u}S}^{\prime 2}}{r_{p}^{\prime 2}}-1}\nonumber\\
&-\frac{1}{2}\frac{s}{r_{p}^{\prime }}\sqrt{\dfrac{r_{Sch\ddot{u}E}^{\prime }-r_{p}^{\prime }}{r_{Sch\ddot{u}E}^{\prime }+r_{p}^{\prime }}}-\frac{1}{2}\frac{s}{r_{p}^{\prime }}\sqrt{\dfrac{r_{Sch\ddot{u}S}^{\prime }-r_{p}^{\prime }}{r_{Sch\ddot{u}S}^{\prime }+r_{p}^{\prime }}}.\label{fischuS-prime}
\end{eqnarray}
Let us now compute the polar angle $\varphi_{Sch\ddot{u}S}$ at
which the lower trajectory photon penetrates inside the
Sch\"{u}cking sphere. Since $\varphi$ decreases when $r$ varies
from $r_{Sch\ddot{u}S}$ to $r_p$ and also when $r$ varies from
$r_p$ to $r_{Sch\ddot{u}E}$, then
\begin{equation}
\varphi _{Sch\ddot{u}S}= \varphi _{Sch\ddot{u}E}+\int_{r_{p}}^{r( t_{Sch\ddot{u}S}) }\left\vert \frac{d\varphi }{dr}\right\vert dr+\int_{r_{p}}^{r( t_{Sch\ddot{u}E}) }\left\vert \frac{d\varphi }{dr}\right\vert dr.
\end{equation}
To the linear order in the ratio $s/r_p$, one gets for $\varphi
_{Sch\ddot{u}S}$ \cite{Sch-strong}
\begin{eqnarray}
 \varphi _{Sch\ddot{u}S}&\simeq& \varphi _{Sch\ddot{u}E}+\pi -\arcsin \left(\dfrac{r_{P}}{r_{Sch\ddot{u}E}}\right)-\arcsin \left(\dfrac{r_{P}}{r_{Sch\ddot{u}S}}\right)\nonumber\\
 &&+\frac{M}{8\pi }\frac{1}{r_{Sch\ddot{u}E}}\sqrt{\dfrac{r_{Sch\ddot{u}E}^{2}}{r_{p}^{2}}-1}+\frac{M}{8\pi }\frac{1}{r_{Sch\ddot{u}S}}\sqrt{\dfrac{r_{Sch\ddot{u}S}^{2}}{r_{p}^{2}}-1}\nonumber\\
 &&+\frac{M}{8\pi }\frac{1}{r_{p}}\sqrt{\dfrac{r_{Sch\ddot{u}E}-r_{p}}{r_{Sch\ddot{u}E}+r_{p}}}+\frac{M}{8\pi }\frac{1}{r_{p}}\sqrt{\dfrac{r_{Sch\ddot{u}S}-r_{p}}{r_{Sch\ddot{u}S}+r_{p}}}.\label{fischuS}
\end{eqnarray}

\item [step 3:] Determination of $t_S^\prime$ and $t_S$

One can now compute $\varphi _{S}^{\prime }$
\begin{equation}
\varphi _{S}^{\prime }=\varphi _{Sch\ddot{u}S}^{\prime }-\gamma _{FS}^{\prime }+ \arcsin\left(\dfrac{\chi _{Sch\ddot{u}}}{\chi_{L,S}}\sin \gamma _{FS}^{\prime }\right),\label{fiS-prime}
\end{equation}
where $\varphi _{Sch\ddot{u}S}^{\prime }$ is given by
(\ref{fischuS-prime}) and where $\gamma _{FS}^{\prime }$ the
smaller physical angle between the un-oriented direction of the
photon and the direction towards the lens as the photon penetrates
inside the Sch\"{u}cking sphere from the external side:
\begin{equation}
\gamma _{FS}^{\prime }=\arctan\!\left(\left \vert \chi
_{Sch\ddot{u}}\dfrac{\dot{\varphi}_{Sch\ddot{u}S}^\prime}{\dot{\chi}_{Sch\ddot{u}S}^\prime}\right
\vert\right)\!= \arctan\!\left(\!-\chi
_{Sch\ddot{u}}\dfrac{\dot{\varphi}_{Sch\ddot{u}S}^\prime}{\dot{\chi}_{Sch\ddot{u}S}^\prime}\right)\!,\label{gammaFS-prime}
\end{equation}
where ${\dot{\varphi}_{Sch\ddot{u}S}^\prime}$ and
${\dot{\chi}_{Sch\ddot{u}S}^\prime}$ are given respectively by
\begin{equation}
\dot{\varphi}_{Sch\ddot{u}S}^{\prime }=\dfrac{r_{p}^{\prime }}{r_{Sch\ddot{u}S}^{\prime 2}\sqrt{B(r_{p}^{\prime })}}=\dfrac{r_{p}^{\prime }}{a^{2}(t_{Sch\ddot{u}S}^{\prime })\chi _{Sch\ddot{u}}^{2}\sqrt{B(r_{p}^{\prime })}},
\end{equation}
and
\begin{eqnarray}
 \dot{\chi}_{Sch\ddot{u}S}^\prime&=&-\frac{1}{a(t_{Sch\ddot{u}S}^{\prime })B_{Sch\ddot{u}}(t_{Sch\ddot{u}S}^{\prime })}\sqrt{1-\frac{r_{p}^{\prime 2}}{a^{2}(t_{Sch\ddot{u}S}^{\prime })\chi _{Sch\ddot{u}}^{2}}\frac{\,B_{Sch\ddot{u}}(t_{Sch\ddot{u}S}^{\prime })}{B(r_{p}^{\prime })}} \nonumber\\
 &&-\frac{C_{Sch\ddot{u}}(t_{Sch\ddot{u}S}^{\prime })}{a(t_{Sch\ddot{u}S}^{\prime })}\frac{1}{B_{Sch\ddot{u}}(t_{Sch\ddot{u}S}^{\prime })},\label{chipointSchuS-prime}
\end{eqnarray}
with
\begin{equation}
 C_{Sch\ddot{u}}(t_{Sch\ddot{u}S}^{\prime
})=\sqrt{\dfrac{A}{a(t_{Sch\ddot{u}S}^{\prime
})}\chi_{Sch\ddot{u}}^2+\dfrac{\Lambda}{3}a^2(t_{Sch\ddot{u}S}^{\prime
})\chi_{Sch\ddot{u}}^2}
\end{equation}
and
\begin{equation}
 B_{Sch\ddot{u}}(t_{Sch\ddot{u}S}^{\prime
})=1-\dfrac{A}{a(t_{Sch\ddot{u}S}^{\prime
})}\chi_{Sch\ddot{u}}^2-\dfrac{\Lambda}{3}{a^2(t_{Sch\ddot{u}S}^{\prime
})}\chi_{Sch\ddot{u}}^2.
\end{equation}
To obtain (\ref{chipointSchuS-prime}) we have used the Jacobian of the coordinate transformation $(T,r)\rightarrow (t,\chi)$, (\ref{Inv-Jacob}), together with the expressions of $\dot{T}_{Sch\ddot{u}S}^{\prime }$ and $\dot{r}_{Sch\ddot{u}S}^{\prime }$
\begin{equation}
 \dot{T}_{Sch\ddot{u}S}^{\prime }=
\dfrac{1}{B(t_{Sch\ddot{u}S}^\prime)}, \quad
\dot{r}_{Sch\ddot{u}S}^{\prime }=-\sqrt{1-\dfrac{r_p^{\prime
2}}{r_{Sch\ddot{u}S}^{\prime
2}}\dfrac{B(t_{Sch\ddot{u}S}^\prime)}{B(r_p^\prime)}}.
\end{equation}

$\chi _{L,S}$ is the geodesic distance between the source and the lens, which can be accurately approximated by
\begin{equation}
\chi _{L,S}\simeq\chi _{S}-\chi _{L}.
\end{equation}

In the same manner, once $\varphi_{Sch\ddot{u}S}$ determined, one
can compute the polar angle $\varphi_S$ corresponding to the
source by a relation analogous to
(\ref{fiS-prime})
\begin{equation}
\varphi _{S}=\varphi
_{Sch\ddot{u}S}+\gamma _{FS}- \arcsin\left(\dfrac{\chi
_{Sch\ddot{u}}}{\chi_{L,S}}\sin \gamma _{FS}\right),
\end{equation}

%

where $\varphi _{Sch\ddot{u}S}$ is given by (\ref{fischuS}) and
where $\gamma _{FS}$ is the smaller physical angle between the
un-oriented direction of the lower trajectory photon and the
direction towards the lens as the photon penetrates inside the
Sch\"{u}cking sphere from the external side:
\begin{equation}
\gamma _{FS}= \arctan\left(\left\vert\chi
_{Sch\ddot{u}}\dfrac{\dot{\varphi}_{Sch\ddot{u}S}}{\dot{\chi}_{Sch\ddot{u}S}}\right\vert\right)=
\arctan\left(\chi
_{Sch\ddot{u}}\dfrac{\dot{\varphi}_{Sch\ddot{u}S}}{\dot{\chi}_{Sch\ddot{u}S}}\right).\label{gammaFS}
\end{equation}
Using
\begin{eqnarray}
\dot{\chi}_{Sch\ddot{u}S}&=&\dfrac{-1}{a(t_{Sch\ddot{u}S})B_{Sch\ddot{u}S}(t_{Sch\ddot{u}S})}
\sqrt{1-\dfrac{r_{p}^{2}}{r_{Sch\ddot{u}S}^{2}}\frac{B_{Sch\ddot{u}S}(t_{Sch\ddot{u}S})}{B(r_{p})}}\nonumber \\
&&-\frac{C_{Sch\ddot{u}S}(t_{Sch\ddot{u}S})}{a(t_{Sch\ddot{u}S})B_{Sch\ddot{u}S}(t_{Sch\ddot{u}S})},
\end{eqnarray}
obtained by using the Jacobian of the coordinate transformation
$(T,r)\rightarrow (t,\chi)$ and
\begin{equation}
\dot{\varphi}_{Sch\ddot{u}S}=-\frac{r_{p}}{a^{2}(t_{Sch\ddot{u}S})\chi
_{Sch\ddot{u}}^{2}\sqrt{B(r_{p})}},
\end{equation}
then
\begin{equation}
 \mbox{\fontsize{8}{13}\selectfont $ \displaystyle{\gamma
_{FS}=\arctan \left(
\dfrac{r_{p}B_{Sch\ddot{u}S}(t_{Sch\ddot{u}S})}{r_{Sch\ddot{u}S}\sqrt{B(r_{p})}\left(
C_{Sch\ddot{u}S}(t_{Sch\ddot{u}S})+\sqrt{1-\dfrac{r_{p}^{2}}{r_{Sch\ddot{u}S}^{2}}\dfrac{B_{Sch\ddot{u}S}(t_{Sch\ddot{u}S})}{B(r_{p})}}\right)
}\right)}$}.
\end{equation}
For a given $M$, one obtains in turn $t_{Sch\ddot{u}E}^{\prime }$, $t_{Sch\ddot{u}S}^{\prime }$, $\varphi _{Sch\ddot{u}S}^{\prime }$, $\varphi _{S}^{\prime }$, $t_{Sch\ddot{u}E}$, $t_{Sch\ddot{u}S}$, $\varphi _{Sch\ddot{u}S}$ and $\varphi _{S}$.

In general $\varphi _{S}^{\prime }\neq \varphi _{S}$. To achieve
$\varphi _{S}^{\prime }= \varphi _{S}$, which corresponds to the
fact that the upper and the lower trajectories photons are emitted
by the same source, we have to adjust $M$, i.e., we have to vary
$M$ until the equality $\varphi _{S}^{\prime }= \varphi _{S}$ is
satisfied. We end up with values of $M$, $t_{Sch\ddot{u}E}^{\prime
}$, $t_{Sch\ddot{u}S}^{\prime }$, $\varphi _{Sch\ddot{u}S}^{\prime
}$, $t_{Sch\ddot{u}E}$, $t_{Sch\ddot{u}S}$, $\varphi
_{Sch\ddot{u}S}$ and $\varphi _{S}^{\prime }=\varphi _{S}$.

We are now in a position to determine $t_S-t_S^\prime$.

Using once again some elementary Euclidean geometry, similar to
that used to obtain $\chi _{Sch\ddot{u}E,E}^\prime$
(\ref{chiEprime}), one obtains for the geodesic distance $\chi
_{Sch\ddot{u}S,S}^\prime$ between the source $S$ and the photon of
the upper trajectory as it crosses the Sch\"{u}cking sphere in the
half space containing the source:
\begin{equation}
\chi _{S,Sch\ddot{u}S}^{\prime }=\sqrt{\chi _{L,S}^{2}+\chi _{Sch\ddot{u}}^{2}-2\chi _{L,S}\chi _{Sch\ddot{u}}\cos \left( \varphi _{Sch\ddot{u}S}^{\prime }-\varphi _{S}\right) }.
\end{equation}
Proceeding in the same manner, we get for the geodesic distance $\chi _{Sch\ddot{u}S,S}$ between the source $S$ and the photon of the lower trajectory as it crosses the Sch\"{u}cking sphere in the half space containing the source
\begin{equation}
\chi _{Sch\ddot{u}S,S}\medskip =\sqrt{\chi _{L,S}^{2}+\chi
_{Sch\ddot{u}}^{2}-2\chi _{L,S}\chi _{Sch\ddot{u}}\cos \left(
\varphi _{Sch\ddot{u}S}-\varphi _{S}\right) }.
\end{equation}

Making use of the approximations
\begin{equation}
\cos x \simeq 1-x^2/2\qquad \mbox{and} \qquad \sqrt{1+x}\simeq 1+ x/2,
\end{equation}
valid for $\left\vert x\right\vert \ll 1$, one gets
\begin{eqnarray}
\chi _{Sch\ddot{u}S,S}^{\prime }&\simeq& \left( \medskip \chi _{L,S}-\chi _{Sch\ddot{u}}\right) -\dfrac{\left( \varphi _{Sch\ddot{u}S}^{\prime }-\varphi _{S}\right) ^{2}}{2\left( \chi _{L,S}^{-1}-\chi _{Sch\ddot{u}}^{-1}\right) },\\
\chi _{Sch\ddot{u}S,S}&\simeq& \left( \medskip \chi _{L,S}-\chi _{Sch\ddot{u}}\right) -\dfrac{\left( \varphi _{Sch\ddot{u}S}-\varphi _{S}\right) ^{2}}{2\left( \chi _{L,S}^{-1}-\chi _{Sch\ddot{u}}^{-1}\right) }.
\end{eqnarray}
Then
\begin{equation}
\chi _{Sch\ddot{u}S,S}^{\prime }-\chi _{Sch\ddot{u}S,S}\simeq
\dfrac{\left( \varphi _{Sch\ddot{u}S}-\varphi _{S}\right)
^{2}-\left( \varphi _{Sch\ddot{u}S}^{\prime }-\varphi _{S}\right)
^{2}}{2\left( \chi _{L,S}^{-1}-\chi _{Sch\ddot{u}}^{-1}\right)
}.\label{chiS-Sprime}
\end{equation}
On the other hand, equation (\ref{chi(t)}) with the final
condition $\chi(t_{Sch\ddot{u}S}^{\prime })=0$ gives
\begin{equation}
0-\chi
_{Sch\ddot{u}S,S}^\prime=-\int_{t_{S}^\prime}^{t_{Sch\ddot{u}S}^\prime}\dfrac{dt}{a(t)},\label{chiSprime}
\end{equation}
In a similar manner, from (\ref{chi(t)}) with the final
condition $\chi(t_{Sch\ddot{u}S})=0$, one gets
\begin{equation}
0-\chi
_{Sch\ddot{u}S,S}=-\int_{t_{S}}^{t_{Sch\ddot{u}S}}\dfrac{dt}{a(t)}.\label{chiS}
\end{equation}
From (\ref{chiSprime}) and (\ref{chiS}), one deduces an expression
for $\chi _{Sch\ddot{u}S,S}-\chi _{Sch\ddot{u}S,S}^{\prime }$
\begin{eqnarray}
\chi _{Sch\ddot{u}S,S}-\chi _{Sch\ddot{u}S,S}^{\prime }&= & \int_{t_{S}}^{t_{Sch\ddot{u}S}}\dfrac{dt}{a(t)}-\int_{t_{S}^{\prime }}^{t_{Sch\ddot{u}S}^{\prime }}\dfrac{dt}{a(t)}\nonumber\\
&=&\int_{t_{S}}^{t_{S}^{\prime }}\dfrac{dt}{a(t)}+\int_{t_{S}^{\prime }}^{t_{Sch\ddot{u}S}}\dfrac{dt}{a(t)}-\int_{t_{S}^{\prime }}^{t_{Sch\ddot{u}S}^{\prime }}\dfrac{dt}{a(t)}\nonumber\\
 &=& \int_{t_{S}}^{t_{S}^{\prime }}\dfrac{dt}{a(t)}+\int_{t_{Sch\ddot{u}S}^{\prime }}^{t_{Sch\ddot{u}S}}\dfrac{dt}{a(t)}.
\end{eqnarray}

But since $a(t)$ varies significantly only on time intervals of
cosmological nature
\begin{eqnarray}
\int_{t_{S}}^{t_{S}^{\prime }}\dfrac{dt}{a(t)}&\simeq \frac{t_{S}^{\prime }-t_{S}}{a(t_{S}^{\prime })}
\int_{t_{Sch\ddot{u}S}^{\prime }}^{t_{Sch\ddot{u}S}}\dfrac{dt}{a(t)}&\simeq  \frac{t_{Sch\ddot{u}S}-t_{Sch\ddot{u}S}^{\prime }}{a(t_{Sch\ddot{u}S}^{\prime })}.
\end{eqnarray}
Then
\begin{equation}
\chi _{Sch\ddot{u}S,S}-\chi _{Sch\ddot{u}S,S}^{\prime }\simeq
\frac{t_{S}-t_{S}^{\prime }}{a(t_{S}^{\prime
})}+\frac{t_{Sch\ddot{u}S}-t_{Sch\ddot{u}S}^{\prime
}}{a(t_{Sch\ddot{u}S}^{\prime })}.\label{chiS-Sprime1}
\end{equation}
Equating the right hand sides of (\ref{chiS-Sprime}) and
(\ref{chiS-Sprime1}), one arrives at
\begin{equation}
 \frac{t_{S}-t_{S}^{\prime }}{a(t_{S}^{\prime })}+\frac{t_{Sch\ddot{u}S}-t_{Sch\ddot{u}S}^{\prime }}{a(t_{Sch\ddot{u}S}^{\prime })}\simeq \dfrac{\left( \varphi _{Sch\ddot{u}S}-\varphi _{S}\right) ^{2}-\left( \varphi _{Sch\ddot{u}S}^{\prime }-\varphi _{S}\right) ^{2}}{2\left( \chi _{L,S}^{-1}-\chi _{Sch\ddot{u}}^{-1}\right) }.
\end{equation}
Then one deduces an expression for $t_{S}-t_{S}^{\prime }$
\begin{equation}
 t_{S}-t_{S}^{\prime }\simeq a(t_{S}^{\prime })\left[\frac{t_{Sch\ddot{u}S}^{\prime }-t_{Sch\ddot{u}S}}{a(t_{Sch\ddot{u}S}^{\prime })}+\dfrac{\left( \varphi _{Sch\ddot{u}S}-\varphi _{S}\right) ^{2}-\left( \varphi _{Sch\ddot{u}S}^{\prime }-\varphi _{S}\right) ^{2}}{2\left( \chi _{L,S}^{-1}-\chi _{Sch\ddot{u}}^{-1}\right) }\right].
\end{equation}
\end{itemize}

Hereafter are displayed the results for
$\varphi_S(=\varphi_S^\prime)$ the deflexion angle, $M$ the mass
of the cluster of galaxies (the lens) and $\Delta t=
t_{S}-t_{S}^{\prime }$, the time delay, in the case of the lensed
quasar SDSS J1004+4112 where \cite{Inada,Oguri,Ota}. '$\pm 0$'
stands for the central value, '$+$' and '$-$' stand respectively
for the upper and the lower experimental limits.
\begin{equation}
\begin{array}{cclcccl}
\alpha ^{\prime }&=&5^{\prime\prime} \pm 10\%,&&   \alpha &=&10^{\prime\prime} \pm 10\%\\
z_{L}&=&0.68,&& z_{S}&=&1.734.
\end{array}
\end{equation}
The cluster mass $M$ comes from a fitting: the angles $\varphi_S$
and $\varphi_S^\prime$ are calculated as a function of $M$ which
is varied until the equality $\varphi_S^\prime=\varphi_S$ is
satisfied.

\pagebreak
For the cosmological $\Lambda$, we take the experimentally favored
$\Lambda\!=\!0.77\! \cdot\!  3 $ am$ ^{-2} \pm 20\% $.
We have also considered the case without cosmological constant
$\Lambda\!=\!0$ in table \ref{tab4}.

\begin{table}[!h]
\begin{minipage}{\linewidth}
      \centering
\caption{Upper limit value of $\Lambda$:\quad$\Lambda=0.77 \cdot  3 $
am$ ^{-2} + 20\% $ } \label{tab1}
      \begin{tabular}{cccccccccc}
         \hline
$\alpha _{E}\pm 10\%$ & $\pm 0$ & $\pm 0$  & $\pm 0$ & + & + & + & $-$ & $-$ & $-$ \\
$\alpha _{E}^{\prime}\pm 10\%$ & $\pm 0$ & + & $-$ & $\pm 0$ & + & $-$ & $\pm 0$ & + & $-$ \\
$$-$\varphi_{S}[^{\prime \prime}] $ & $9.03$ & $8.13$ & $9.94$ & $10.84$ & $9.94$ & ${\bf 11.74} $ & $7.23$ & ${\bf 6.32} $ & $8.13$\\
$M\,[10^{13}M_{\odot}] $ &  $1.80$ & $1.98$ & $1.62$ & $1.98$ & ${\bf 2.18} $ & $1.78$ & $1.62$ & $1.78$ & ${\bf 1.46} $ \\
$\Delta t[\mbox{years}] $ &  $9.76$ & $9.18~$ & $10.25$ & $12.35$ & $11.81$ & ${\bf 12.77} $ & $7.38$ & ${\bf 6.74~} $ & $7.91~$ \\
         \hline
      \end{tabular}
\end{minipage}
\end{table}
\vspace{-0.3cm}
\begin{table}[!h]
\begin{minipage}{\linewidth}
      \centering
\caption{Central value of $\Lambda$:\quad$\Lambda=0.77 \cdot  3 $ am$
^{-2} $ } \label{tab2}
      \begin{tabular}{cccccccccc}
         \hline
$\alpha _{E}\pm 10\%$ & $\pm 0$ & $\pm 0$  & $\pm 0$ & + & + & + & $-$ & $-$ & $-$ \\
$\alpha _{E}^{\prime}\pm 10\%$ & $\pm 0$ & + & $-$ & $\pm 0$ & + & $-$ & $\pm 0$ & + & $-$ \\
$$-$\varphi_{S}[^{\prime \prime}] $ & $9.97$ & $8.98$ & $10.97$ & $11.97$ & $10.97$ & $ {\bf12.97}$ & $7.98$ & $ {\bf6.98} $ & $8.98$ \\
$M\,[10^{13}M_{\odot}] $ &  $1.82$ & $2.00$ & $1.64$ & $2.00$ & ${\bf2.21}$ & $1.80$ & $1.64$ & $1.80$ & ${\bf1.48}$ \\
$\Delta t[\mbox{years}] $ &  $9.72$ & $9.14$ & $10.19$ & $12.28$ & $11.76$ & ${\bf12.68}$ & $7.35$ & ${\bf6.73}$ & $7.87$ \\
         \hline
      \end{tabular}
\end{minipage}
\end{table}
\vspace{-0.3cm}
\begin{table}[!h]
\begin{minipage}{\linewidth}
      \centering
\caption{Lower limit value of $\Lambda$:\quad$\Lambda=0.77 \cdot  3 $
am$ ^{-2} - 20\% $ } \label{tab3}
      \begin{tabular}{cccccccccc}
         \hline
$\alpha _{E}\pm 10\%$ & $\pm 0$ & $\pm 0$  & $\pm 0$ & + & + & + & $-$ & $-$ & $-$ \\
$\alpha _{E}^{\prime}\pm 10\%$ & $\pm 0$ & + & $-$ & $\pm 0$ & + & $-$ & $\pm 0$ & + & $-$ \\
$$-$\varphi_{S}[^{\prime \prime}] $ & $10.57$ & $9.51$ & $11.63$ & $12.68$ & $11.63$ & $ {\bf13.74}$ & $8.46$ & ${\bf7.40}$ & $9.51$\\
$M\,[10^{13}M_{\odot}] $ &  $1.80$ & $1.98$ & $1.62$ & $1.98$ & ${\bf 2.18} $ & $1.79$ & $1.62$ & $1.79$ & ${\bf 1.46} $ \\
$\Delta t[\mbox{years}] $ &  $9.53~$ & $8.97~$ & $9.98$ & $12.03$ & $11.53$ & ${\bf 12.41} $ & $7.21~$ & ${\bf 6.60~} $ & $7.72~$ \\
         \hline
      \end{tabular}
\end{minipage}
\end{table}
\vspace{-0.3cm}
\begin{table}[!h]
\begin{minipage}{\linewidth}
      \centering
\caption{ $\Lambda =0$} \label{tab4}
      \begin{tabular}{cccccccccc}
         \hline
$\alpha _{E}\pm 10\%$ & $\pm 0$ & $\pm 0$  & $\pm 0$ & + & + & + & $-$ & $-$ & $-$ \\
$\alpha _{E}^{\prime}\pm 10\%$ & $\pm 0$ & + & $-$ & $\pm 0$ & + & $-$ & $\pm 0$ & + & $-$ \\
$$-$\varphi_{S}[^{\prime \prime}] $ & $11.86$ & $10.67$ & $13.05$ & $14.23$ & $13.05$ & $ {\bf15.42}$ & $9.49$ & ${\bf8.30}$ & $10.67$\\
$M\,[10^{13}M_{\odot}] $ &  $1.68$ & $1.84$ & $1.51$ & $1.84$ & ${\bf 2.03} $ & $1.66$ & $1.51$ & $1.66$ & ${\bf 1.36} $ \\
$\Delta t[\mbox{years}] $ &  $8.70~$ & $8.20~$ & $9.10$ & $10.97$ & $10.53$ & ${\bf 11.30} $ & $6.59~$ & ${\bf 6.04~} $ & $7.05~$ \\
         \hline
      \end{tabular}
\end{minipage}
\end{table}
\newpage \section{Conclusion} In this paper, we have computed the time
delay caused by a spherical mass, a cluster of galaxies, in
presence of a cosmological constant $\Lambda$ using the
Einstein-Straus solution, which is the appropriate framework for
taking into account the precession of the observer and the effect
of the other masses of the universe in the form of a homogeneous
isotropic dust, the observer being taken comoving  with the dust.
We have applied our results to the lensed quasar SDSS J1004+4112.
We have computed the time delay between the images C and D of the
quasar, which are the most aligned images with the lens, and
obtained results compatible with the lower bound given by
Fohlmeister \cite{Fohlmeister}. Our predictions of the time delay
range from 6 to 13 years.

It is worthwhile to compare our results with those of previous
computations performed by Sch\"{u}cker and Zaimen
\cite{sch-zaimen} and by Kawano and Oguri \cite{Kawano}.
Sch\"{u}cker and Zaimen computed the time delay in the framework
of the Kottler solution and obtained predictions ranging from 13
to 28 years, using however a different mass of the lens. On the other hand,
Kawano and Oguri obtained a time delay of 10 years.

In addition to the hypothesis of sphericity we have made in our
calculations an additional assumption: we have supposed implicitly
that the photons don't penetrate the mass distribution region,
since we have used only the exterior Kottler solution inside the
vacuole. It is then worthwhile to repeat the calculations taking
into account that the photons can penetrate the interior of the
mass distribution, where an interior Kottler solution must be used
\cite{Sch-interior}.

\section*{Acknowledgment} This work was supported by Le
Minist\`{e}re de l'Enseignement Sup\'{e}rieur et de la Recherche
Scientifique of Algeria under grant D00920090096.



\begin{thebibliography}{99}
\bibitem{rindler-ishak} W. Rindler and M. Ishak, {\it The Contribution of the Cosmological Constant to the Relativistic Bending of Light Revisited}, Phys. Rev. {\bf D76} 043006 [arXiv:0709.2948 [astro-ph]] (2007).
\bibitem{sereno1} M. Sereno, {\it On the influence of the cosmological constant on gravitational lensing in small systems.}, Phys. Rev. {\bf D77}, 043004 [arXiv:0711.1802 [astro-ph]] (2008)
\bibitem{Sereno2} M. Sereno, {\it The role of Lambda in the cosmological lens equation.} [arXiv:0807.5123 [astro-ph]]
\bibitem{Sch-Lambda-lens} T. Sch\"{u}cker, {\it Cosmological constant and lensing.} [arXiv:0712.1559 [astro-ph]], Gen. Relativ. Gravit. DOI:10.1007/s10714-008-0652-2
\bibitem{sch-proceed} T. Sch\"{u}cker, {\it Strong lensing with positive cosmological constant.} [arXiv:0805.1630 [astro-ph]].Moriond Proceedings Cosmology (2008)
\bibitem{Miraghaei} H. Miraghaei, M. Nouri-Zonoz, {\it Classical tests of general relativity in the Newtonian limit of Schwarzschild-de-Sitter spacetime.} [arXiv:0810.2006 [gr-qc]]
\bibitem{Kantowski} R. Kantowski, B. Chen and X. Dai, {\it Gravitational lensing corrections in flat $\Lambda$CDM cosmology.} [arXiv:0909.3308v1 [astro-ph.CO]]
\bibitem{Khriplovich} I.B. Khriplovich, A.A. Pomeransky, {\it Does cosmological term influence gravitational lensing?} [arXiv:0801.1764 [gr-qc]]
\bibitem{park} M. Park, {\it Rigorous approach to the gravitational lensing.} [arXiv:0804.4331 [astro-ph]]
\bibitem{Gibbons} G.W. Gibbons, C.M. Warnick, M.C. Werner, {\it Light-bending in Schwarzschild-de-Sitter: projective geometry of the optical metric.} [arXiv:0808.3074 [gr-qc]]
\bibitem{Simpson} F. Simpson, J.A. Peacock, A.F. Heavens, {\it On lensing by a cosmological constant.} [arXiv:0809.1819 [astro-ph]]
\bibitem{Sch-strong} T. Sch\"{u}cker, {\it Strong lensing in the Einstein-Straus solution.} Gen Relativ Gravit (2009) 41:1595-1610 DOI:10.1007/s10714-008-0731-4
\bibitem{Einstein-Straus} A. Einstein, E.G. Straus, {\it The influence of the expansion of space on the gravitation fields surrounding the individual star.} Rev. Mod. Phys. {\bf 17}, 120 (1945), {\bf 18}, 148 (1946)
\bibitem{Schucking} E. Sch\"{u}cking, {\it Das Schwarzschildsche Linienelement und die Expansion des Weltalls.} Z. Phys. {\bf 137}, 595 (1954)
\bibitem{Ishak-Rind-Doss} M. Ishak, W. Rindler, J. Dossett, {\it More on Lensing by a Cosmological Constant.} [arXiv:0810.4956 [astro-ph]]
\bibitem{Balbinot} R. Balbinot, R. Bergamini, A. Comastri, {\it Solution of the Einstein-Straus problem with a $Lambda$ term.} Phys.
Rev. {\bf D38}, 2415 (1988)
\bibitem{sch-zaimen} T. Sch\"{u}cker,
N. Zaimen, {\it Cosmological constant and time delay.} A\&A {\bf
484}, 103 [arXiv:0801.3776 [astro-ph]] (2008)
\bibitem{Inada} N. Inada et al., [SDSS Collaboration], {\it A Gravitationally Lensed Quasar with Quadruple Images Separated by 14.62 Arcseconds,} Nature {\bf 426}, 810 [arXiv:astro-ph/0312427] (2003)
\bibitem{Oguri} M. Oguri et al., [SDSS Collaboration], {\it Observations and Theoretical Implications of the Large Separation Lensed Quasar SDSS J1004+4112.} Astrophys. J. {\bf 605}, 78 [arXiv:astro-ph/0312429] (2004)
\bibitem{Ota} N. Ota et al., {\it Chandra Observations of SDSS J1004+4112: Constraints on the Lensing Cluster and Anomalous X-Ray Flux Ratios of the Quadruply Imaged Quasar}. Astrophys. J. {\bf 647}, 215 [arXiv:astroph/0601700] (2006)
\bibitem{Fohlmeister} J. Fohlmeister et al., {\it The Rewards of Patience: An 822 Day Time Delay in the Gravitational Lens SDSS J1004+4112.} [arXiv:0710.1634
[astro-ph]] (2007)
\bibitem{Kawano} Y. Kawano and M. Oguri, {\it Time delays for the Giant Quadruple Lensed SDSS J1004+4112: Prospects for Determining the Densityn Profile of the Lensing Cluster} [arXiv:0601149v1 [astro-ph]] (2006)
\bibitem{Sch-interior} T. Sch\"{u}cker, {\it Lensing in an interior Kottler
solution} [arXiv:0903.2940 [astro-ph]] (2009)
\end{thebibliography}
\end{document}